\documentclass[prb,twocolumn,amsmath,amssymb,floatfix]{revtex4}

\usepackage{graphicx}% Include figure files

\def\gammav{{\mbox{\boldmath{$\gamma$}}}}
\def\sigmav{{\mbox{\boldmath{$\sigma$}}}}

\usepackage{amssymb}
\usepackage{color}
\usepackage{epsfig}

\begin{document}

\def\gammav{{\mbox{\boldmath{$\gamma$}}}}
\def\sigmav{{\mbox{\boldmath{$\sigma$}}}}
%\noindent{\bf {\large Discrete versus continuous wires on quantum networks}}
\title{Discrete versus continuous wires on quantum networks}
%\vspace{6mm}

%\hspace{21mm}{\bf Amnon Aharony and Ora Entin-Wohlman}

\author{Amnon Aharony}

\altaffiliation{Also emeritus, Tel Aviv University.}

%\hspace{21mm}{\it Department of Physics and the Ilse Katz Center for
%Meso- and Nano-Scale Science and Technology, Ben Gurion
%University, Beer Sheva 84105, ISRAEL}

\affiliation{Department of Physics and the Ilse Katz Center for
Meso- and Nano-Scale Science and Technology, Ben Gurion
University, Beer Sheva 84105, ISRAEL}

\author{Ora Entin-Wohlman}

\altaffiliation{Also emeritus, Tel Aviv University.}

\affiliation{Department of Physics and the Ilse Katz Center for
Meso- and Nano-Scale Science and Technology, Ben Gurion
University, Beer Sheva 84105, ISRAEL}

%\end{widetext}
%%% ----------------------------------------------------------------------
\date{\today}

\begin{abstract}
Mesoscopic systems and large molecules are often modeled by graphs of one-dimensional
wires, connected at vertices.
 %In 1981, P. G. de Gennes discussed superconductivity on such networks,
%by solving  the linearized Landau-Ginzburg equations and using the
%so-called Kirchoff (or Neumann) matching conditions at the
%vertices. 
In this paper we discuss the solutions of the Schr\"odinger
equation on such graphs, which have been named ``quantum networks".
Such solutions are needed for finding the energy spectrum of single electrons on such finite systems
or for finding the transmission of electrons between leads which connect such systems to reservoirs.
Specifically, we compare two common approaches. In the ``continuum" approach, one solves the one-dimensional
Schr\"odinger equation on 
each continuous wire, and then uses the Neumann-Kirchoff-de Gennes matching conditions
at the vertices. Alternatively, one replaces each wire by
a finite number of ``atoms", and then uses the
tight binding model for the solution. Here we show that these approaches
cannot generally give the same results, except for special energies. Even in the limit of vanishing lattice constant, the
two approaches coincide only if the tight binding parameters obey
very special relations. The different consequences of the two
approaches are demonstrated via the example of a T-shaped scatterer.
\end{abstract}
%\pacs{71.70.Ej, 72.25.-b, 73.23.Ad}

\maketitle

\noindent {\bf I. Introduction}
\vspace{4mm}

The quantum mechanics of electrons on networks, made of
one-dimensional (1D) wires which are connected at vertices, has
been the subject of active research for a long time. This research
probably started with the study of organic molecules, in 1953.
\cite{RS,G}  However, it gained a large momentum in 1981, when P.
G. de Gennes considered the linearized Landau-Ginzburg equations
on networks, in connection with percolating
superconductors.\cite{dG1,dG2} This work was later extended by
Alexander.\cite{SA} In the last 25 years, quantum networks have
been used as theoretical models of mesoscopic
systems, which have become experimentally available due to the
advancements in microfabrication. In these models, the
1D wires represent limits of three dimensional wave guides, which
are assumed to be sufficiently narrow to justify keeping only the lowest
transverse mode. Quantum networks  are also used to model
novel molecular devices, where each wire consists of discrete
atoms. 

In this paper we compare several approaches for solving the single electron
Schr\"odinger equation on such networks. 
Basically, this solution  consists of two stages. First, one solves the equation on
each wire. Denoting the vertices at the ends of the wire by
$\alpha$ and $\beta$, one can express the solution inside the
wire, $\psi_{\alpha\beta}(x_{\alpha\beta})$ (where
$x_{\alpha\beta}=0$ at vertex $\alpha$ and
$x_{\alpha\beta}=L_{\alpha\beta}$ at vertex $\beta$) in terms of
the wave functions at its ends, $\Psi_\alpha$ and $\Psi_\beta$. At
the next stage, one must employ some matching conditions at each
vertex. Using these matching conditions, and eliminating the wave
functions inside each wire, one ends up with a set of discrete
linear equations for the wave functions at the vertices,
\begin{align}
M_{\alpha\alpha}\Psi_\alpha+\sum_\beta M_{\alpha\beta}\Psi_\beta=0,
\label{eq1}
\end{align}
where the coefficients $M_{\alpha\beta}$ are determined by the 1D
solution on each wire and by the matching conditions. Obviously,
the detailed solution for the network depends on the latter.

There have been several approaches to the issue of the matching
conditions. Some of the earlier papers in mesoscopics used the
scattering approach.\cite{BIA,GIA,B} For a vertex with ${\cal
N}_\alpha$ wires, this approach constructs an ${\cal
N}_\alpha\times {\cal N}_\alpha$ unitary scattering
matrix,\cite{Shapiro}  which relates the amplitudes of the ${\cal
N}_\alpha$ incoming (plane) waves to the ${\cal N}_\alpha$
outgoing waves. Given this matrix, one can solve for the
amplitudes of the two waves in each wire, and thus find a full
solution for the wave functions on the network. For the examples
treated in these early papers, the scattering matrix was
parametrized arbitrarily. To find explicit values for the elements
of the scattering matrix, one must solve a more ``microscopic"
model, as done e.g. in Refs. \onlinecite{Kowal} and
\onlinecite{Kottos2}.

A widely used second approach follows de Gennes,\cite{dG1,dG2} and
uses the so called Kirchoff (or Neumann) matching condition at the
vertices.\cite{Xia,Deo,Kottos1,ak,Vidal,Bercioux,Bercioux1} Given the
continuous solutions $\psi_{\alpha\beta}(x_{\alpha\beta})$ on each
of the ${\cal N}_\alpha$ wires which meet at vertex $\alpha$, one
requires continuity of the wave function,
$\psi_{\alpha\beta}(0)\equiv \Psi_\alpha$, and also the condition
\begin{align}
\sum_{\beta=1}^{{\cal N}_\alpha}\frac{\partial
\psi_{\alpha\beta}}{\partial x_{\alpha\beta}}\Big
|_{{}_{x_{\alpha\beta}=0}}=0,\label{kirch}
\end{align}
where the sum is over the ${\cal N}_\alpha$ wires which meet at vertex $\alpha$.
Since the outgoing probability current on wire $(\alpha\beta)$ is
proportional to
\begin{align}
I_{\alpha\beta}=\Im[(\psi_{\alpha\beta})^\ast(\partial \psi_{\alpha\beta}/\partial x_{\alpha\beta})],\label{curr}
\end{align}
Eq. (\ref{kirch}) supplies a sufficient condition for the
conservation of the probability at this vertex.

However, Eq. (\ref{kirch}) does not represent a {\it necessary} condition for the conservation of current. A more
general condition is given by
\begin{align}
\sum_\beta\frac{\partial \psi_{\alpha\beta}}{\partial x_{\alpha\beta}}\Big |_{{}_{x_{\alpha\beta}=0}}
=\lambda_\alpha\Psi_\alpha,\label{BC}
\end{align}
with any real $\lambda_\alpha$.\cite{avron,Kot3,TM,TB} Multiplying
both sides of this equation by $(\Psi_\alpha)^\ast$
yields a real number, which does not contribute to the current in
Eq. (\ref{curr}). For a vertex with two wires, $\lambda_\alpha$
represents the strength of a delta-function potential at the vertex.\cite{TB} An infinite value
of $\lambda_\alpha$ in Eq. (\ref{BC}) implies that
$\Psi_\alpha=0$, which is known as the Dirichlet matching
condition. In this limit, there is no probability current through
this vertex.  Thus, finite values of $\lambda_\alpha$ interpolate
between the Neumann and the Dirichlet limits. However, for real
networks one still needs to  derive $\lambda_\alpha$
from some microscopic equations.

An alternative approach, which avoids some of these ambiguities,
is based on the {\it tight binding model}, as done e.g. in Ref.
\onlinecite{Kowal}. In this approach, the system is described by a
set of localized wave functions, associated with discrete
``atomic" sites $\{n\}$. This approach is often used for
molecules, where it is equivalent to the method of linear
combination of atomic orbitals (LCAO). It is also used for the
calculation of energy bands in periodic lattices.\cite{MA}  The discrete Schr\"odinger equations for the
amplitudes $\{\psi_n\}$ of the ``atomic" stationary wave functions for an energy $\epsilon$ are then given by
\begin{align}
(\epsilon-\epsilon_n)\psi_n=-\sum_m J_{nm}\psi_m.
\label{TB}
\end{align}
Here, $\epsilon_n$ and $-J_{nm}$ are the diagonal and the
off-diagonal matrix elements of the Hamiltonian, in the basis of
the atomic wave functions. For real molecules or crystals, these
matrix elements can be calculated from the atomic wave functions
and from the microscopic potentials seen by the electron.\cite{MA}
The tight binding equations have also been applied to a variety of specific
mesoscopic quantum networks, where the ``atoms" represent quantum
dots; see e. g. Refs. \onlinecite{Kowal}, \onlinecite{VMD,AA1,AA2,Filter} and
references therein.

For 1D chains, the results of the discrete tight-binding model approach those of the
continuous Schr\"odinger equation
when one sends the lattice
constant of the former to zero (see below). However, in this paper we show that generally this
limit is problematic for more complex networks. Section II
presents the derivation of Eq. (\ref{eq1}) for continuous wires
and for discrete wires (described by the tight binding Hamiltonian), and contains a critical comparison
between them. Section III demonstrates the difference between the two approaches for the simplest example of a T-shaped scatterer,
 and Sec. IV
contains our conclusions.

\vspace{8mm}

\noindent {\bf II. The equations for the vertex wave functions}
\vspace{4mm}

As stated above, applying the Neumann matching conditions to the wave functions describing the continuous wires
 yields Eq. (\ref{eq1}). Here we first
rederive this equation for the more general matching condition (\ref{BC}),
then derive a similar equation from the solution of the discrete model,
and finally compare the two resulting equations.

\vspace{15mm}

\noindent {\bf A. Continuous graphs}
\vspace{3mm}

For simplicity, we follow much of the literature and assume that
there is no potential energy within the wires. We start by solving the continuous 1D Schr\"odinger
equation for a free particle on each 1D wire. For a given energy, $\epsilon$,
the solutions are
linear combinations of $e^{ikx}$ and $e^{-ikx}$, with
$\epsilon=\hbar^2k^2/(2m)$. Here we used the simplifying
assumption that the effective mass $m$ is the same on all the
wires (in general one could have a different wave vector $k$ for each wire, for the same energy $\epsilon$).
 We next express the solution $\psi_{\alpha\beta}$ on the
wire $(\alpha\beta)$ in terms of the wave functions at its two
ends,\cite{avron,Vidal} which we denote by $\Psi_\alpha$ and $\Psi_\beta$:
\begin{align}
\psi_{\alpha\beta}(x_{\alpha\beta})=\Psi_\alpha \frac{\sin[k(L_{\alpha\beta}
-x_{\alpha\beta})]}{\sin(kL_{\alpha\beta})}+\Psi_\beta\frac{\sin(kx_{\alpha\beta})}{\sin(kL_{\alpha\beta})}.\label{psi}
\end{align}
Substituting these expressions into the matching condition
(\ref{BC}) then yields
\begin{align}
\sum_\beta k\bigl
[-\Psi_\alpha\cot(kL_{\alpha\beta})+\Psi_\beta/\sin(kL_{\alpha\beta})\Bigr
] =\lambda_\alpha\Psi_\alpha.
\end{align}
Generally $k$ could have different values for different wires. Assuming the same $k$ for all wires, this can be rewritten in the form of Eq. (\ref{eq1}), with
\begin{align}
M_{\alpha\alpha}=\frac{\lambda_\alpha}{k}+\sum_\beta
\cot(kL_{\alpha\beta}),\ \ M_{\alpha\beta}
=-\frac{1}{\sin(kL_{\alpha\beta})}.\label{Mab}
\end{align}
Indeed, $\lambda_\alpha \rightarrow \infty$ implies that $M_{\alpha\alpha}\rightarrow \infty$,
and therefore $\Psi_\alpha=0$, as in the Dirichlet limit.
The solution of the set of linear equations (\ref{eq1})  then either yields ratios among the $\Psi$'s at a given energy $\epsilon$ (for a scattering problem) or gives the set of spectrum of allowed $\epsilon$'s (for an eigenvalue problem).

\vspace{5mm}
\noindent {\bf B. The tight binding model}
\vspace{3mm}

We now replace the $(\alpha\beta)$ wire by a discrete chain of
``atoms". For simplicity, we assume that all the wires have the
same lattice constant $a$. The wire $(\alpha\beta)$ now has
$N_{\alpha\beta}+1$ ``atoms", at sites $x_{\alpha\beta}=na$, where
$n=0,1,2,...,N_{\alpha\beta}$. The length of the wire is then given by
$L_{\alpha\beta}=aN_{\alpha\beta}$. Assuming no
potential energy within each wire also allows us to set the
diagonal site energies $\epsilon_n$ within each wire equal to
zero. For simplicity we also assume uniform nearest-neighbor ``hopping" matrix
elements, $J_{n,n\pm 1}=J$. For $0<n<N_{\alpha\beta}$, the tight
binding equation (\ref{TB}) becomes
\begin{align}
\epsilon\psi_{\alpha\beta}(n)=-J[\psi_{\alpha\beta}(n-1)+\psi_{\alpha\beta}(n+1)],\label{TB1D}
\end{align}
with solutions $e^{ikna}$ or $e^{-ikna}$ and with the dispersion
relation \begin{align} \epsilon=-2J\cos(ka). \label{dis}
\end{align}
(Again, if different wires have different $J$'s or different $a$'s then each wire will have its own $k$, for the same $\epsilon$).

Fixing the wave functions at the ends of the wire,
$\psi_{\alpha\beta}(0)=\Psi_\alpha$ and
$\psi_{\alpha\beta}(N_{\alpha\beta})=\Psi_\beta$, the solution
along the chain $(\alpha\beta)$ becomes
\begin{align}
\psi_{\alpha\beta}(n)=\frac{\sin[ka(N_{\alpha\beta}-n)]}{\sin(kaN_{\alpha\beta})}\Psi_\alpha
+\frac{\sin(kan)}{\sin(kaN_{\alpha\beta})}\Psi_\beta.\label{psiTB}
\end{align}
As might be expected, this solution is just a discrete version of
Eq. (\ref{psi}), but with the modified dispersion relation
(\ref{dis}).

We next write down the tight binding equation (\ref{TB}) for the vertex site
$\alpha$, where we do allow for a non-zero site energy
$\epsilon_\alpha$:
\begin{align}
(\epsilon-\epsilon_\alpha)\Psi_\alpha=-J\sum_\beta
\psi_{\alpha\beta}(n=1).\label{TBver}
\end{align}
Substituting Eq. (\ref{psiTB}), and using Eq. (\ref{dis}) (with the same $k$ on all wires), this
again yields Eq. (\ref{eq1}), and $M_{\alpha\beta}$ is again given
by Eq. (\ref{Mab}). However, the diagonal coefficient is
different:
\begin{align}
&M_{\alpha\alpha}=\sum_\beta[\cot(kaN_{\alpha\beta})-\cot(ka)]\nonumber\\
&+2\cot(ka)+\epsilon_\alpha/[J\sin(ka)].\label{MTB}
\end{align}
Note that in some sense Eq. (\ref{eq1}) has the same form as the
discrete Schr\"odinger equation  (\ref{TB}) for the vertices of the network.
However, unlike the $J_{nm}$'s and $\epsilon_n$'s of that equation, the coefficients $M_{\alpha\alpha}$ and $M_{\alpha\beta}$
do depend on the energy $\epsilon$ which characterizes the solution.
Note also that these renormalized coefficients
reduce back to those in Eq. (\ref{TB}) when all the wires reduce
to single bonds, $N_{\alpha\beta}=1$.

\vspace{5mm}
\noindent {\bf C. Comparison}
\vspace{3mm}

Comparing Eq. (\ref{MTB}) with Eq.
(\ref{Mab}), we see that the two equations would become identical
only when
\begin{align} \lambda_\alpha/k=\epsilon_\alpha/[J\sin(ka)]-({\cal
N}_\alpha-2)\cot(ka). \label{lam}\end{align} 
(Again: the terms which depend on $ka$ 
should be replaced by sums of such terms with different $k$'s if the wires have different parameters).
The on-site energy on the vertex
$\epsilon_\alpha$ represents a diagonal matrix element of the
Hamiltonian, and therefore it should not depend on the energy $\epsilon$  which represents the specific solution for the Schr\"odinger
equation. Therefore, Eq. (\ref{lam}) implies that the ``equivalent" continuum Hamiltonian would have parameters
$\lambda_\alpha$ which generally depend on $\epsilon$. However, the
$\lambda_\alpha$'s represent the matching conditions, and these are
also characteristic of the system and not of its solutions.
Therefore, in general one cannot choose energy-independent vertex parameters for which the two approaches  yield the same
solutions everywhere.

Interestingly, the difference between the two approaches is largest when
$\cot(ka)$ is large, namely near the band edges.
Usually, divergences near the 1D band edges are associated with van Hove
singularities, which cause the divergence of the density of states. For this reason, the use of
tight binding equations is often restricted to the vicinity of the band
center, i. e. $ka \sim \pi/2$. However, even in this region the two models
become identical only {\it at} the band center $\epsilon=0$ and not at any other energy. Furthermore, although one can choose
the electron energy (i. e. the Fermi energy) in a scattering calculation, where
the spectrum is continuous, this choice may be limited in a calculation of the
whole (discrete) spectrum of a finite network, or in a calculation which may
involve several bands (or discrete energies) with gaps between them.

It is interesting to consider this comparison for a vertex on a
linear chain, when ${\cal N}_\alpha=2$. Even in this case, the
parameters $\lambda_\alpha$ or $\epsilon_\alpha$ would have to
depend on energy, except for the trivial case
$\lambda_\alpha=\epsilon_\alpha=0$. The latter condition simply
implies that this vertex is equivalent to any other point along
the wire on which it ``sits". Although appropriate in some cases,
this cannot be generally valid. For example, this does not suffice
to describe the configuration of two wires which meet at a finite angle.
\cite{Bercioux} Such a
break in the straight line should cause scattering even if the wire itself 
is free of any scattering centers. To describe this scattering, one would 
need to introduce non-zero values for
$\epsilon_\alpha$ and/or $\lambda_\alpha$.
At this stage, we are not aware of a systematic study of this issue.

A priori, one might have thought that the two approaches should
coincide when the lattice constant in the tight binding model
approaches zero. Indeed, as noted above, the two models yield the
same answers within each (uniform) linear chain. In the limit of
vanishing $a$, Eq. (\ref{TB1D}) can be written as
\begin{align}
(\epsilon+2J)\psi_{\alpha\beta}(n)&=-J[\psi_{\alpha\beta}(n-1)
+\psi_{\alpha\beta}(n+1)\nonumber\\
&-2\psi_{\alpha\beta}(n)] \approx
-Ja^2\frac{\partial^2\psi_{\alpha\beta}}{\partial x^2}\Big
|_{_{x=na}}.
\end{align}
This reproduces the continuum Schr\"odinger equation, provided we
move the zero of energy to the bottom of the band, $-2J$, and
identify the effective mass $m$ via $\hbar^2/(2m)=Ja^2$,
requiring that either $m\rightarrow \infty$ or $J \rightarrow \infty$ and $a \rightarrow 0$, such that
$Ja^2$ is kept finite.

We next turn to the Schr\"odinger equation at the vertex, Eq. (\ref{TBver}).  In the limit of small $a$, we
rewrite this equation as
\begin{align}
&(\epsilon-\epsilon_\alpha+{\cal N}_\alpha J)\Psi_\alpha\nonumber\\
&=-J\sum_\beta[\psi_{\alpha\beta}(1)-\psi_{\alpha\beta}(0)]
\approx -Ja\sum_\beta\frac{\partial \psi_{\alpha\beta}}{\partial
x_{\alpha\beta}}\Big |_{{}_{x_{\alpha\beta}=0}}.
\end{align}
For a finite effective mass $m$, $(\epsilon+2J)\sim Ja^2k^2$ remains finite when $a
\rightarrow 0$, and this equation becomes equivalent to Eq. (\ref{BC}) with
\begin{align}
\lambda_\alpha \sim[\epsilon_\alpha-({\cal N}_\alpha-2)J]/(Ja).\label{lim}
\end{align}
In fact, this equation also follows from Eq. (\ref{lam}) in the
limit of vanishing $a$. For ${\cal N}_\alpha\ne 2$, the two models
coincide in this limit only if either $\lambda_\alpha \rightarrow
\infty$ (i.e. the Dirichlet matching condition which implies $\Psi_\alpha=0$)
or $\epsilon_\alpha \sim ({\cal N}_\alpha-2)J \sim 1/a^2 \rightarrow \infty$
 i. e.
\begin{align}
\epsilon_\alpha\approx ({\cal N}_\alpha-2)J+\mu_\alpha/a,\ \
a\rightarrow 0. \label{inf}
\end{align}
The correction term, $\mu_\alpha$, determines $\lambda_\alpha$ via
$\lambda_\alpha=\mu_\alpha/(Ja^2)$. To reproduce the
Kirchoff-Neumann-de Gennes matching conditions (\ref{kirch}), one would need
to have $\mu_\alpha=0$.

At this stage, we have no direct
information on how the tight binding Hamiltonian matrix elements
vary with $a$ for small $a$. In fact, the tight binding model was
constructed to deal with small overlaps between the localized wave
functions, which is no longer valid when $a \rightarrow 0$.
However, it would still be interesting to have explicit
microscopic mathematical calculations of the tight binding matrix elements, and to
check whether indeed they all diverge as $1/a^2$. It would be even more interesting to find out
whether the limiting tight binding matrix elements obey the
relation (\ref{inf}).

An exception arises for a vertex on a linear chain,  ${\cal
N}_\alpha=2$. In this case, Eq. (\ref{lim}) reduces to
$\lambda_\alpha \sim\epsilon_\alpha/(Ja)$. For a finite effective mass on the wires, $Ja^2$ remains finite
when $a\rightarrow 0$, and therefore $\lambda_\alpha
\rightarrow 0$ whenever $\epsilon_\alpha$ is finite, reproducing
the Neumann matching condition.

Some of the papers which use tight binding models to describe
mesoscopic systems actually use $\epsilon_n=0$ for {\it all} the
sites, except for a few resonant sites for which $\epsilon_n$ is
varied by an external gate voltage. As explained, such a variation
does not affect the Neumann matching conditions for ${\cal N}_\alpha=2$ when $a\rightarrow 0$. However, if
such gate voltages are applied to a vertex
with more than two wires, then the parameter $\lambda_\alpha$ for that vertex will increase with decreasing $a$,
crossing over to the Dirichlet condition and to the vanishing of the wave function on that vertex when $a \rightarrow 0$.

\vspace{8mm}

\noindent {\bf III. The T-shape scatterer}
\vspace{4mm}

\begin{figure}[ht]
\begin{center}
\includegraphics[width=7.3 cm]{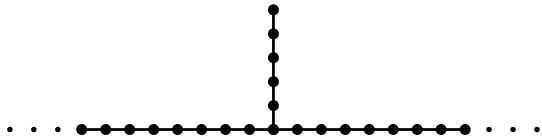}
\end{center}
\caption{A wire with a stub. }\label{1}
\end{figure}

Quantum networks have been studied for a variety of geometries.
Here we demonstrate
the differences between the discrete tight binding and the continuum
models on one of the simplest geometries. 
We calculate the transmission of an electron passing through a
T-shaped device, consisting of a main wire attached to a stub of
length $L$ perpendicular to the wire at the origin $x=0$ (Fig.
\ref{1}).\cite{Deo} Assuming that the electrons come from the left hand side,
we expect a reflected wave on the left and a transmitted wave on the right hand side.

\begin{figure}[ht]
\begin{center}
\includegraphics[width=7.3 cm]{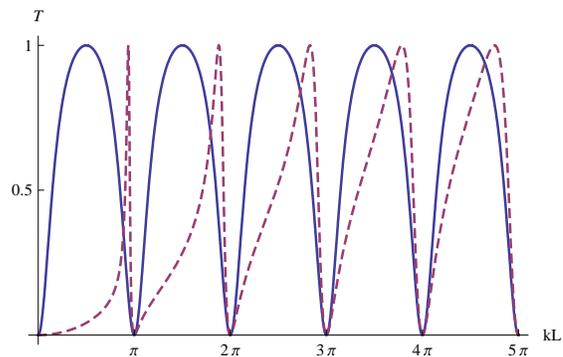}
\end{center}
\caption{Transmission for the continuum stub. The full
(dashed) line represents $\lambda=0$ ($\lambda=3$). }\label{2}
\end{figure}

 In the continuum approach, the wave
function on the main wire is
\begin{align}
\psi_w(x)&=e^{ikx}+r e^{-ikx}, \ \ \ x\le 0,\nonumber\\
\psi_w(x)&=t e^{ikx}, \ \ \ \ \ \ \ \ \ \ \ \ \ \ x \ge 0,\label{stub}
\end{align}
where $r$ and $t$ are the reflection and transmission amplitudes.
Thus, $\psi_w(0)=1+r=t$. From Eq. (\ref{psi}), the wave function
on the stub is given by
\begin{align}
\psi_s(y)=\psi_w(0) \frac{\sin[k(L -y)]}{\sin(kL)},\label{st1}
\end{align}
vanishing at $y=L$. Using Eq. (\ref{BC}) at the origin we find
$t=2/(2+i[\cot(kL)+\lambda/k])$, and therefore the transmission
through the main wire is
\begin{align}
T=|t|^2=4/\bigl (4+[\cot(kL)+\lambda/k]^2\bigr ).\label{TC}
\end{align}
We note that this expression does not reproduce $T=1$ for $L=0$ and $\lambda=0$, probably
due to problems with the order of limits.
For $L>0$, $T$ oscillates with $k$ between 0 (full reflection) and 1
(full transmission). Sample results are shown in Fig. \ref{2}.
Since $T(k)=T(-k)$, we show only results for
$k>0$.  When $\lambda=0$,
$T(k)$ is periodic, with period $\pi/L$. However, for a non-zero
energy-independent $\lambda$ the transmission is no longer periodic, approaching the
zero-$\lambda$ behavior only for large $k$.

We next consider the same model using the discrete equations. With a
finite lattice constant $a$, the wave function is written as in
Eq. (\ref{stub}), with $x$ replaced by $na$. Similarly, we set $L=(N+1)a$ and replace
$y$ in Eq. (\ref{st1}) by $ma$, with $m=0,1,...,N+1$, where $N$
represents the number of bonds on the stub (5 in Fig. \ref{1}), after setting
$\psi_s(N+1)=0$. The tight binding equation at the origin then
yields
\begin{align}
t=2/\bigl (2+i\bigl
[\cot[ka(N+1)]+\epsilon_0/[J\sin(ka)]-\cot(ka)\bigr ]\bigr ),\label{stTB}
\end{align}
and $T=|t|^2$. It is easy to see the replacement given in Eq.
(\ref{lam}). Note that (unlike Eq. (\ref{TC}), Eq. (\ref{stTB}) does reproduce $T=1$ for $N=0$ and $\epsilon_0=0$.

\begin{figure}[ht]
\begin{center}
\includegraphics[width=7.3 cm]{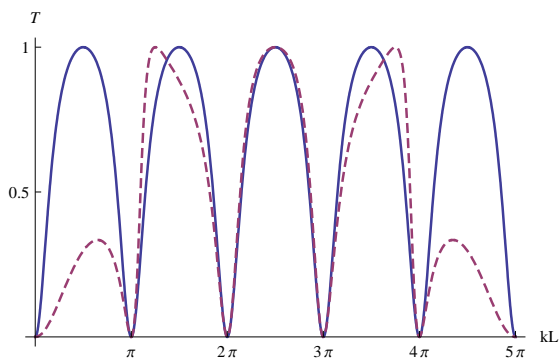}
\end{center}
\caption{Transmission for the continuum stub (full line), with
$\lambda=0$,  and the tight binding stub (dashed line), with
$\epsilon_0=0$, for $N+1=L/a=5$.  }\label{3}
\end{figure}

Figure \ref{3} compares the continuum result with $\lambda=0$
(same as in Fig. \ref{2}) with the tight binding result with
$\epsilon_0=0$, for $a=L/5$. As seen from Eq. (\ref{stTB}), the transmission
is now periodic with the new larger period $\pi/a$ ($=5\pi/L$ for the case drawn
in Fig. \ref{3}). Therefore, we show only the lower half of the
energy band, $0<k<\pi/a$. As expected, the two models differ
mostly near the edges of the energy band,
where $\cot(ka)$ diverges. 
In contrast, the two approaches give very close results
near the center of the band, $\epsilon \sim 0$
or $ka\sim \pi/2$.
However, the difference between the two models increases as one moves away from this region.

\begin{figure}[ht]
\begin{center}
\includegraphics[width=7.3 cm]{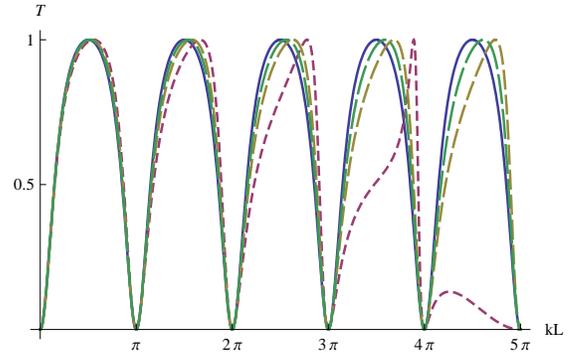}
\end{center}
\caption{Same as Fig. \ref{3}, but with $\epsilon_0/J={\cal N}_0-2=1$.
The dashed curves correspond to
$N+1=5,~10$ and $20$ (increasing dashes). }\label{4}
\end{figure}

The above results obviously change with the parameters $\lambda$
and $\epsilon_0$. For example, Fig. \ref{4} compares  the
continuum model with $\lambda=0$ with the tight binding model with
$\epsilon_0/J=({\cal N}_0-2)=1$, for several values of $a$. As might be anticipated from Eq. (\ref{lam}),
this value of $\epsilon_0$ brings the tight binding
equations close to the continuum ones near the bottom of the band,
i. e. at small $k$. Indeed, Fig. \ref{4} shows almost a
coincidence of the results near $k=0$. This coincidence improves as $N+1=L/a$ increases.
However, the results are
very different near the center of the band, $ka\sim \pi/2$.

\vspace{8mm}

\noindent {\bf IV. Conclusions}
\vspace{4mm}

A detailed comparison between the tight binding ``atomic" approach and the continuum approach to quantum networks reveals that the two models are generally different from each other, and one cannot expect them to yield the same results for all energies. However, the two models agree quantitatively near the center of the 1D band (which describes each of the wires in the network), $\epsilon \sim 0$. The two models are definitely very different near the band edges, where one encounters the van Hove singularities. 
Some of the differences between the models can be removed by adding appropriate site energies to the tight binding model, or by tuning the parameters of the matching conditions at the vertices of the continuous graphs. Sending the lattice constant to zero brings the two models close to each other near the bottom of the band only for very special choices of the site energies.

This paper presented only one simple example of a scattering solution, to demonstrate our main points. In such a scattering solution, where one is free to choose the energy of the scattered electrons near $\epsilon=0$, the difference between the two approaches is small. This freedom does not exist in other situations. Foe example, we have recently followed Bercioux {\it et al.}\cite{Bercioux,Bercioux1} in considering an
infinite chain of diamonds, where the electrons are subject to both the Aharonov-Bohm flux\cite{AB}
and the Rashba spin-orbit interactions.\cite{rashba} With these two effects, 
the various wave functions become 2-component spinors, and the basic hopping terms
become $2\times 2$ unitary matrices, which also contain phase factors. Furthermore, the probability current contains additional terms, due to these phases.
Unlike Bercioux {\it et al.}, who found the spectrum of the  diamond chain within the continuum model,
we solved the same model within the tight binding approach, where each edge of each diamond represented a single bond.\cite{Filter}
We have since repeated this calculation with more ``atoms" within each edge.
Interestingly, the qualitative features of the resulting energy bands in the two approaches are similar. However, the interesting physics (e. g. the spin filtering found in Ref. \onlinecite{Filter}) does involve moving away from the center of the basic 1D band. Therefore, the two approaches do yield different quantitative results.
The continuum and the discrete approaches certainly give different results when one calculates the spectrum of a finite network. In this case, the resulting energies can assume any value, including values which are outside the ``basic" band. In the latter case, the related $k$'s are imaginary.

In the above discussion we have not attempted to judge which approach is ``better". As we mentioned in passing, both approaches are sometimes oversimplified, missing some important physics. In practical situations, it is probably best to use both approaches and compare the results.

\vspace{5mm}
{\bf Acknowledgements.} This paper is dedicated to the memory of
Pierre Gilles de Gennes, with whom we had many illuminating
discussions on many fields in physics. We acknowledge discussions
and correspondence with Y. Imry, U. Smilansky and J. Vidal, and
support from the ISF and from the DIP.

\end{document}